# Photographing a time interval

## Bernhard Rothenstein and Ioan Damian

Politehnnica University of Timisoara, Department of Physics,
Timisoara, Romania
**bernhard_rothenstein@yahoo.com, ijdamian@yahoo.com**

**Abstract** *A method of measuring time intervals by a single observer proposed by Crowell[3] is extended to the more general case when the events separated by the time interval take place at two points characterized by the same y=y' space coordinates. We show that time dilation and time contraction can take place or even an inversion in the time succession can be detected.*

### 1. Introduction

Weinstein[1] shows how a single observer could measure the length of a moving rod by simultaneously detecting the light signals that have left the two ends of the rod at different times. He considers only the case in which the observer is very close to the rod. Mathews[2] presents the same problem in the general case and stresses that because the light signals arise from different distances to from the observer, the measured length of the rod will be very different from that given by the Lorentz contraction. He considers that the measurement procedure is associated with the Doppler Effect as well. We show that the involved observers are concerned with the Doppler Effect only in the case when the frequency and the light source-observer distance are very high.

Crowell[3] shows that this argument can be applied to the observation of a time interval by a single observer. He considers the same simple case as Weinstein[1] did. The purpose of our paper is to generalize the approach proposed by Crowell[3].

### 2. Photographing a time interval

Consider the scenario presented in Figure 1. It involves a rod of proper length $L_0$ at rest in the K'(X'O'Y') reference frame, parallel to the O'X' axis. The space coordinates of its edges are $\mathbf{1'}\left(x_1' = r_1'\cos\theta_1', y_1' = r_1'\sin\theta_1'\right)$ and $\mathbf{2'}\left(x_2' = r_2'\cos\theta_2', y_2' = r_2'\sin\theta_2'\right)$, using both Cartesian and polar coordinates. At a time $t'=0$, an observer $R_0'(0,0)$ (the photographer in K') located at the origin $O'$ receives the light signal that has left edge $\mathbf{1'}$ at a time $t_1' = -\dfrac{r_1'}{c}$ and the light signal that has left edge $\mathbf{2'}$ at a time $t_2' = -\dfrac{r_2'}{c}$. The time $t_1'$ is displayed by a clock $C_1'(x_1', y_1')$, whereas a clock $C_2'(x_2', y_2')$ displays the time $t_2'$. All the clocks of the K'(X'O'Y') reference frame are synchronized following the clock synchronization procedure proposed by Einstein[4]. The events associated with the fact that the light signals have left the edges of the rod are $\mathbf{E_1'}\left(x_1', y_1', t_1'\right)$ and $\mathbf{E_2'}(x_2', y_2', t_2')$ respectively. From Figure 1 we obtain

$$r_1'\cos\theta_1' = r_2'\cos\theta_2' - L_0 \tag{1}$$

$$r_1'\sin\theta_1' = r_2'\sin\theta_2' \tag{2}$$

from which we obtain

$$r_1' = r_2'\sqrt{1 + \left(\frac{L_0}{r_2'}\right)^2 - 2\left(\frac{L_0}{r_2'}\right)\cos\theta_2'} \ . \tag{3}$$

The time separation between the two events as measured in K' is

$$\frac{cT'}{L_0} = \frac{r_2'}{L_0}\left(1 - \sqrt{1 + \left(\frac{L_0}{r_2'}\right)^2 - 2\left(\frac{L_0}{r_2'}\right)\cos\theta_2'}\right) =$$

$$\frac{h}{L_0 \sin\theta_2'}\left(1 - \sqrt{1 + \left(\frac{L_0}{h}\right)^2 \sin^2\theta_2' - \left(\frac{L_0}{h}\right)\sin 2\theta_2'}\right) \qquad (4)$$

In Figure 2 we present the variation of $\frac{cT'}{L_0}$ with $\theta_2'$ for $\beta = \frac{V}{c} = 0.6$ and different values of

$\frac{L_0}{r_2'} = \frac{L_0}{h}\sin\theta_2'$, where $h$ represents the distance from the points where the events take place to the common axes. As we see, depending on the relative position of the two points, where the events take place, the time interval $T'$ can be positive if both points are "outgoing ($r_2' > r_1'$)" but it can be negative as well if both of them are "incoming ($r_2' < r_1'$)" or even equate a zero value when $r_1' = r_2'$ i.e. when $x_1' = -\frac{L_0}{2}$ and $x_2' = \frac{L_0}{2}$. For $\theta_1' = \theta_2' = 0$ and for $\theta_1' = \theta_2' = \pi$ we have $\frac{T'}{L_0} = \pm 1$ because in that case $T' = \frac{L_0}{c}$ the light signals originating from **1'** and **2'** respectively propagate along the line that joins the two events. As involving physical quantities measured in the same inertial reference frame, the derivation of (4) involves special relativity by the constancy of the velocity of light postulate.

Consider the experiment described above from the K(XOY) reference frame. The axes of the two reference frames are parallel to each other, the OX(O'X') axes are common and K' moves with constant velocity $V$ relative to K in the positive direction of the overlapped axes. At the origin of time in the two reference frames ($t=t'=0$) the origins of the two frames are located at the same point in space.

Detected from K the space-time coordinates of the two events defined above are $\mathbf{E}_1\left(x_1 = r_1\cos\theta_1,\ y_1 = r_1\sin\theta_1,\ t_1 = -\frac{r_1}{c}\right)$ and $\mathbf{E}_2\left(x_2 = r_2\cos\theta_2,\ y_2 = r_2\sin\theta_2,\ t_2 = -\frac{r_2}{c}\right)$. In accordance with the Lorentz-Einstein transformation for the space-time coordinates of the same event we have[5]

$$r_1 = r_1'\left(1 - \beta\cos\theta_2'\right) \qquad (5)$$

$$r_2 = r_2'(1 - \beta\cos\theta_2') . \qquad (6)$$

Therefore the two events are separated in K by the time interval $T$ given by

$$T = \frac{r_2}{c} - \frac{r_1}{c} = \gamma\left(\frac{r_2'}{c} - \frac{r_1'}{c}\right) - \gamma\beta\left(\frac{r_2'}{c}\cos\theta_2' - \frac{r_1'}{c}\cos\theta_1'\right) = \gamma T' - \gamma\beta\frac{L_0}{c} \qquad (7)$$

or

$$\frac{T}{T'} = \gamma\left(1 - \frac{\beta\frac{L_0}{r_2'}}{1 - \sqrt{1 + \left(\frac{L_0}{r_2'}\right)^2 - 2\left(\frac{L_0}{r_2'}\right)\cos\theta_2'}}\right) . \qquad (8)$$

We can express (8) as a function of $h$

$$\frac{T}{T^{'}} = \gamma \left( 1 - \frac{\beta \frac{L_0}{h} \sin\theta_2^{'}}{1 - \sqrt{1 + \left(\frac{L_0}{h}\right)^2 \sin^2\theta_2^{'} - \left(\frac{L_0}{h}\right)\sin 2\theta_2^{'}}} \right). \qquad (9)$$

An inspection of (9) shows that the time interval measured in the K reference frame is the result of a "dilation" of the time interval $T^{'}$ measured in the stationary reference frame K' and of an extra term determined by the fact that the two events take place in K' at different points in space.

We can express (9) as a function of the angle $\theta_2$ measured in K, given by[6]

$$\cos\theta_2 = \frac{\cos\theta_2^{'} - \beta}{1 - \beta\cos\theta_2^{'}} \qquad (10)$$

In Figure 3 we present the variation of $\theta_2$ with $\theta_2^{'}$ for different values of $\beta$. This enables us to evaluate the rotations introduced by the detection procedure we have used.

If the points **1'** and **2'** are "very close" to the OX(O'X') axis and both are "outgoing" ($\theta_1^{'} = \theta_2^{'} = 0$) equation (9) becomes

$$\left(\frac{T}{T^{'}}\right) = \gamma\left(1 - \beta\right) \qquad (11)$$

whereas if both points are "incoming" ($\theta_1^{'} = \theta_2^{'} = \pi$) it leads to

$$\frac{T}{T^{'}} = \gamma\left(1 + \beta\right). \qquad (12)$$

In order to illustrate the results obtained above we in Figure 4 we present the variation of $\frac{T}{T^{'}}$ with $\theta_2^{'}$ as predicted by (9), for different values of $\frac{L_0}{h}$ and $\beta = 0.6$.

In the experimental conditions described above the time interval $T^{'}$ is not a proper time interval. There is a particular situation when $T'$=0, presented in Figure 5, when

$$\cos\theta_2^{'} = \frac{L_0}{2r_2^{'}} \qquad (13)$$

and $\mathbf{E}_1^{'}\left(\frac{L_0}{2}, h, -\frac{1}{c}\sqrt{h^2 + \left(\frac{L_0}{2}\right)^2}\right)$ ; $\mathbf{E}_2^{'}\left(-\frac{L_0}{2}, h, -\frac{1}{c}\sqrt{h^2 + \left(\frac{L_0}{2}\right)^2}\right)$. The two events are separated in K by a time interval

$$T = \gamma\frac{V}{c^2}L_0. \qquad (14)$$

As in the case when the two events take place on the overlapped axes OX(O'X').

If photographer $R_0^{'}(0,0)$ takes a snapshot of a single event say $\mathbf{E}^{'}(r^{'}, \theta^{'})$, then he measures a time interval $T^{'} = \frac{r^{'}}{c}$. Detected from K it is

$$T = \gamma T^{'}\left(1 - \beta\cos\theta^{'}\right). \qquad (15)$$

We also have

$$r = \gamma r^{'}(1 - \beta\cos\theta^{'}). \qquad (16)$$

We can consider that $r$ and $r^{'}$ are multiples of the wavelength $(\lambda, \lambda^{'})$ and that T and T' are periods, so that (15) and (16) describe the Doppler Effect in oblique incidence. In that case $T$ and *T'* are proper time intervals[4]. All other cases studied above have nothing in common with the Doppler Effect as relating non-proper time intervals.

### 3. Conclusions

If the time dilation effect in its common use is the result of a comparison between a proper time interval measured in the rest frame of one of the clocks involved and a non-proper time interval measured in a reference frame relative to which the clock moves, the time dilation and time contraction effects revealed above are the result of the two relativistic effects mentioned above. In order to avoid confusion it is advisable, when we speak about the magnitude of a physical quantity, to mention the observer (observers) who measure it, the measuring devices used and when and where he (they) performs the measurement.

### Figure captions

Figure 1. The scenario involves an observer $R_0^{'}(0,0)$ located at the origin O' of its rest frame and an observer $R(0,0)$ located at the origin O of its rest frame K'(X'O'Y'). At the common origin of time in the two frames they receive two light signals which have left at different times two different points characterized by the same distance $h$ to the overlapped axes and separated by the proper length $L_0$ in the K'(X'O'Y') reference frame. What we compare are the time intervals $T$ and $T'$ in K and K' respectively.

Figure 2. We present the variation of $\dfrac{cT'}{L_0}$ with angle $\theta_2^{'}$ for $\beta = 0.6$ and different values of the characteristic parameter $a = \dfrac{L_0}{h}$.

Figure 3. We present the variation of the angle $\theta_2$ with $\theta_2^{'}$ for different values of $\beta$ as predicted by the aberration of light effect.

Figure 4. We present the variation of $\dfrac{T}{T'}$ with $\theta_2^{'}$ for $\beta = 0.6$ and different values of $a = \dfrac{L_0}{h}$.

Figura 5. A scenario in which observer $R_0^{'}(0,0)$ detects two simultaneous events.

**Figures**

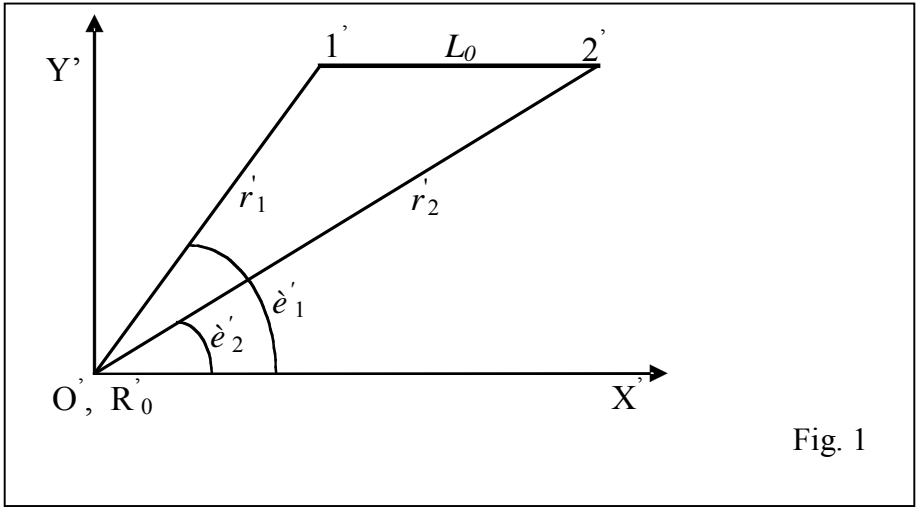

Fig. 1

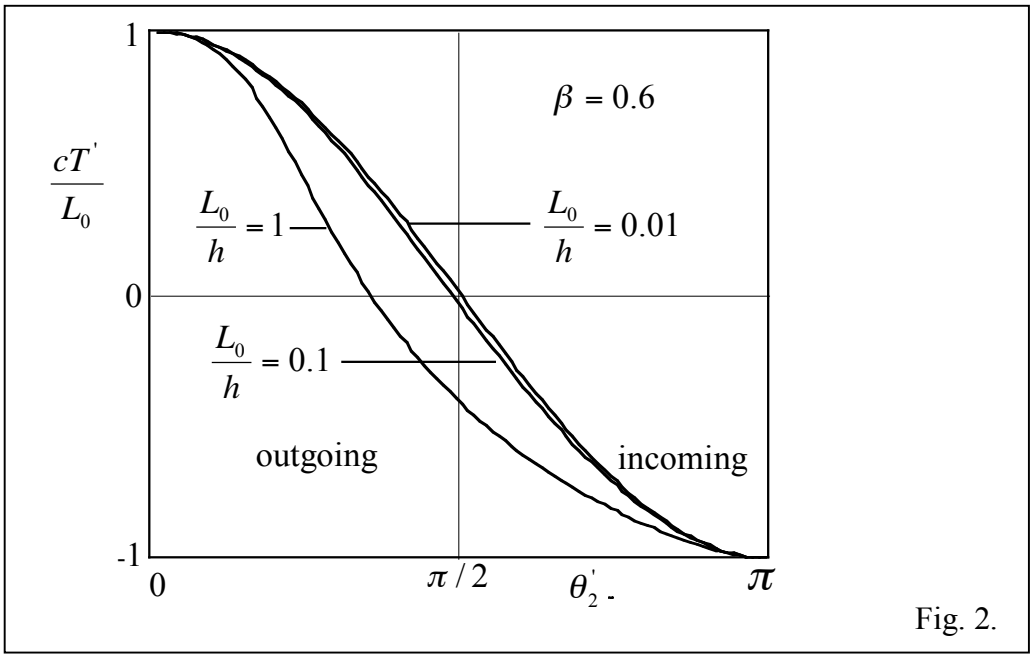

Fig. 2.

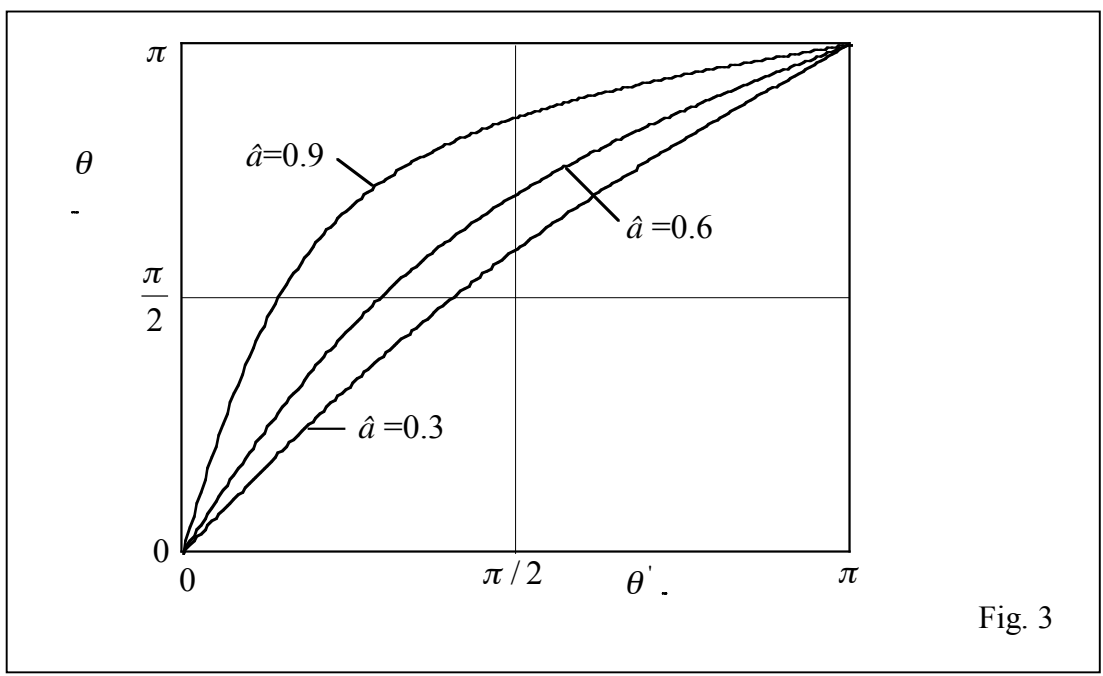

Fig. 3

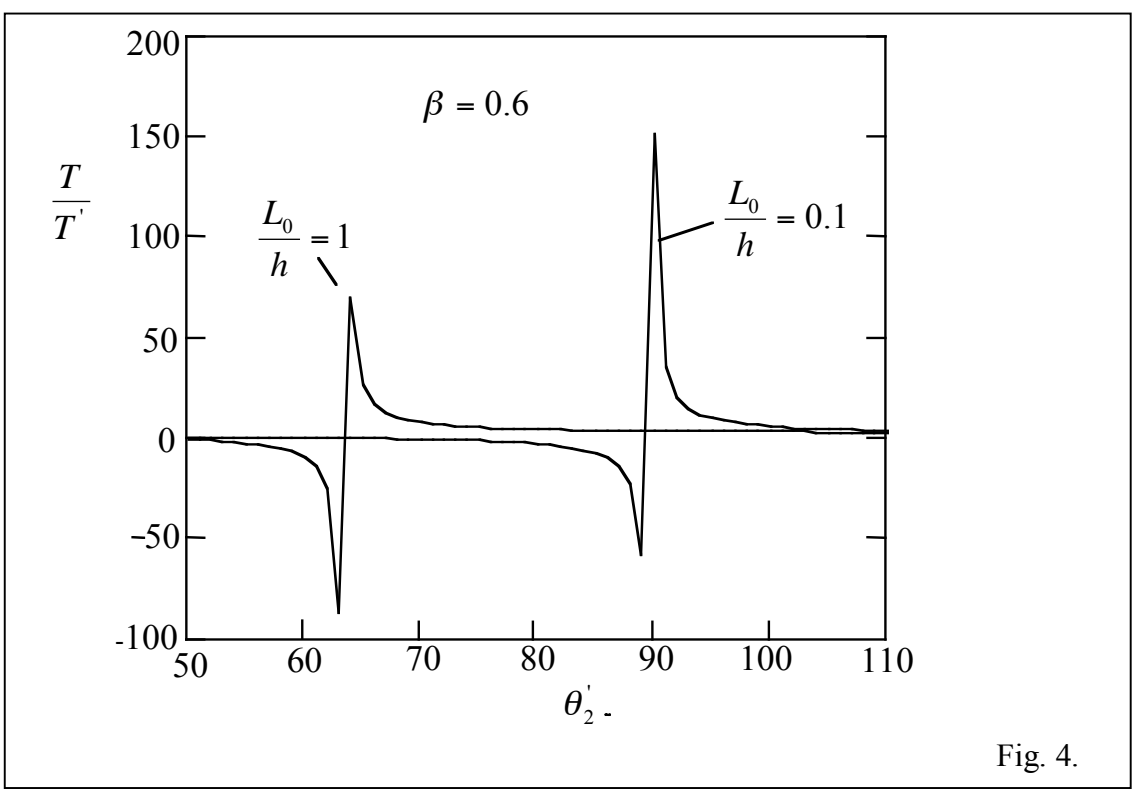

Fig. 4.

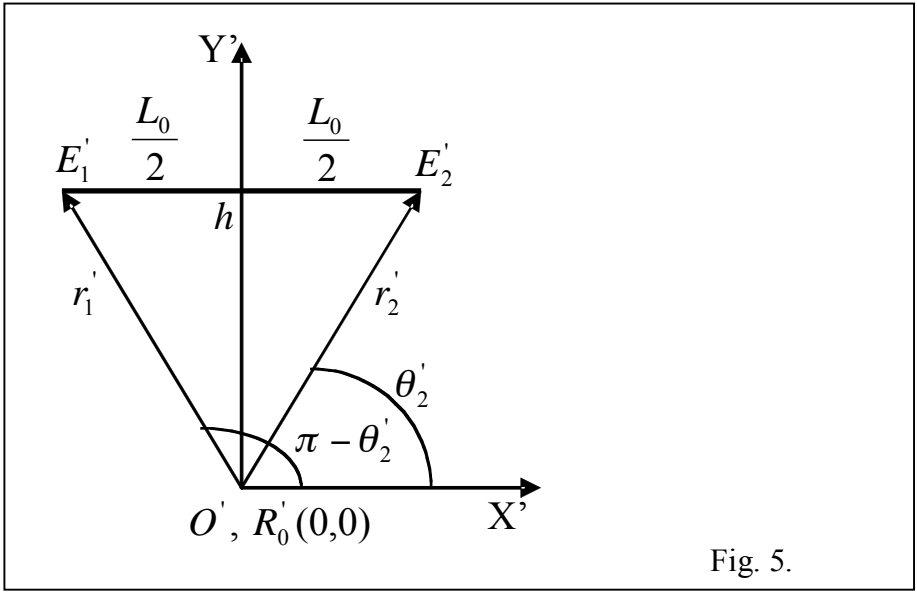

Fig. 5.

7